\documentclass{egpubl}
\usepackage{pg2019}

\usepackage{mathtools}
\usepackage{amsmath}
\usepackage{amssymb}
\usepackage{amsfonts}
\usepackage{amsopn}
\usepackage{textcomp}
\usepackage{xfrac}
\usepackage{bbm}
\usepackage{xcolor}
\usepackage{overpic}
\usepackage{subfig}
\usepackage{wrapfig}
\usepackage[ruled,vlined,linesnumbered]{algorithm2e}
\usepackage{multirow}
\usepackage{enumitem}

\usepackage{color}
\definecolor{green}{rgb}{0, 0.5, 0}
\definecolor{orange}{rgb}{0.8, 0.6, 0.2}
\definecolor{red}{rgb}{1.0, 0.0, 0.0}
\definecolor{teal}{rgb}{0.0, 0.4, 0.4}
\definecolor{purple}{rgb}{0.65,0,0.65}
\definecolor{saffron}{rgb}{0.95,0.75,0.2}
\definecolor{turquoise}{rgb}{0.0,0.5,0.5}

\usepackage{overpic}
\usepackage{currfile} 

\definecolor{lightgray}{rgb}{0.6, 0.6, 0.6}

\usepackage[normalem]{ulem}

\renewcommand{\paragraph}[1]{\textbf{#1.}}

\newcommand{\hidecomment}[1]{}

\usepackage{xspace}

\SpecialIssuePaper         

\usepackage[T1]{fontenc}
\usepackage{dfadobe}

\usepackage{cite}  
\BibtexOrBiblatex
\electronicVersion
\PrintedOrElectronic

\usepackage{egweblnk}

\title[Active Scene Understanding via Online Semantic Reconstruction]%
      {Active Scene Understanding via Online Semantic Reconstruction}

\author[L. Zheng, C. Zhu, J. Zhang, H, Zhao, H. Huang, M. Nie{\ss}ner \& K. Xu]
{\parbox{\textwidth}{\centering Lintao Zheng$^{1}$, Chenyang Zhu$^{1\dagger}$, Jiazhao Zhang$^{1}$, Hang Zhao$^{1}$, Hui Huang$^{2,3}$, Matthias Niessner$^{4}$
        and Kai Xu$^{1}$\thanks{Corresponding authors: Chenyang Zhu (chenyang.chandler.zhu@gmail .com) and Kai Xu (kevin.kai.xu@gmail.com)}
        }
        \\
{\parbox{\textwidth}{\centering $^1$National University of Defense Technology, China\\
         $^2$Shenzhen University, China\\
         $^3$Shenzhen Institutes of Advanced Technology, China\\
       $^4$ Technical University of Munich, Germany
       }
}
}

\begin{document}

\teaser{
 \includegraphics[width=\linewidth]{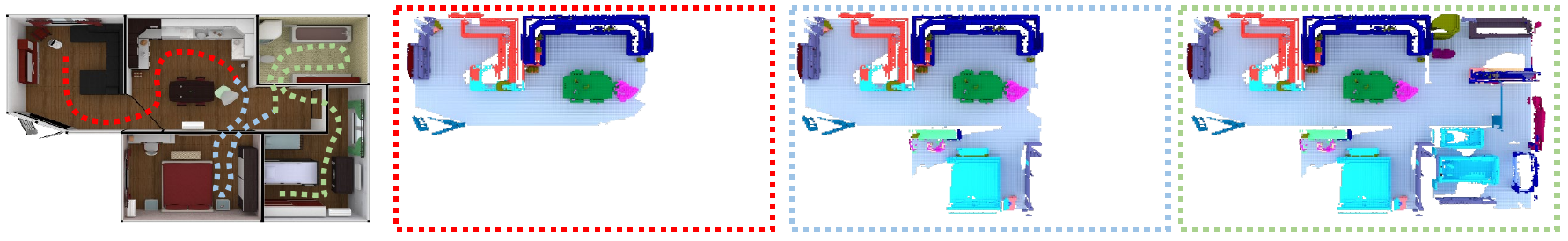}
 \centering
  \caption{We introduce a method for robot-operated active semantic understanding of unknown indoor scenes,
based on online RGBD reconstruction with semantic segmentation. Our approach performs online volumetric RGBD reconstruction, on which real-time voxel-based semantic labeling is conducted. The robot is guided by the requirement of fast online segmentation with minimal scanning effort. The image to the left shows the robot paths computed by our method and the corresponding scene parsing results (correspondence indicated by color) are shown to the right.}
\label{fig:teaser}
}

\maketitle
\begin{abstract}
We propose a novel approach to robot-operated active understanding of unknown indoor scenes,
based on online RGBD reconstruction with semantic segmentation.
In our method, the exploratory robot scanning is both driven by
and targeting at the recognition and segmentation of semantic objects from the scene.
Our algorithm is built on top of a volumetric depth fusion framework 
and performs real-time voxel-based semantic labeling over the online reconstructed volume.
The robot is guided by an online estimated discrete viewing score field (VSF)
parameterized over the 3D space of 2D location and azimuth rotation.
VSF stores for each grid the score of the corresponding view, which measures how much it reduces the uncertainty (entropy) of both geometric reconstruction and semantic labeling.
Based on VSF, we select the next best views (NBV) as the target for each time step.
We then jointly optimize the traverse path and camera trajectory
between two adjacent NBVs, through maximizing the integral viewing score (information gain)
along path and trajectory.
Through extensive evaluation, we show that our method achieves efficient and accurate
online scene parsing during exploratory scanning. 

\end{abstract} 

\section{Introduction}
With the wide availability of commodity RGBD sensors and the boosting of 3D deep learning techniques, 3D scene understanding on RGBD data has been emerging as a core problem of 3D vision and gained much attention from both graphics and vision community lately~\cite{song2015sun,gupta2015indoor,naseer2019indoor}. 
The majority of existing works pursues offline, passive analysis, in which scene understanding, encompassing object detection and/or segmentation, is conducted over already acquired RGBD sequences or their 3D reconstruction.
Unfortunately, this often greatly limits scene understanding performance, since data acquisition is decoupled from the respective scene understanding algorithms; i.e., in some regions, additional observations might be taken in order to make reliable semantic predictions. 

Online scene understanding is a different paradigm in which acquisition and analysis are intertwined~\cite{Xu15,liu2018,ye2018active}: while scene analysis is conducted online based on the progressively acquired scene data, scene scanning, on the other hand, is driven by the requirement of efficient scene understanding. Such a coupled solution fits well for robot-operated autonomous scene understanding.
Here, the robot actively selects scanning views and traversing paths to cover the regions which may best facilitate scene parsing, with a minimum traversing and scanning effort. 

Online scene understanding can be performed either directly over online acquired RGBD sequences or based on online RGBD reconstruction. Most recent works usually adopt the former due to the deep-learning-friendly representation of RGBD images~\cite{gupta2015indoor}. However, 3D object segmentation should best be performed over the 3D reconstruction of scene geometry which facilitates 3D spatial and structural reasoning~\cite{Zhang2014,Xu15}. Modern real-time RGBD reconstruction usually adopts the volumetric depth fusion approach~\cite{Newcombe2011,Izadi2011,Niessner2013,Whelan2015,Dai2017}, where the depth images acquired in real time are registered and fused into a volumetirc representation of scene geomtery, i.e., Truncated Signed Distance Field (TSDF)~\cite{Curless96}. Volumetric representation is well-suited for 3D feature learning based on deep neural networks~\cite{Wu15}.

Inspired by the recent work of semantic scene segmentation with volumetric representation~\cite{hou20183d}, we propose a method of active scene understanding based on online RGBD reconstruction with volumetric segmentation.
Based on the online reconstructed TSDF volume,
our method leverages a deep neural network to perform real-time voxel-based semantic labeling. The network contains a 2D feature extraction module used for extracting 2D features from multi-view RGB images as well as an incremental 3D feature aggregation module specifically designed for real-time inference. The 3D feature fusion and spatial reasoning based on the online updated volume lead to reliable online semantic segmentation.

The robot scanning is guided by an online estimated discrete viewing score field (VSF) parameterized in the 3D view space of 2D location and azimuth rotation. VSF stores for each view a score measuring how much it reduces the uncertainty (entropy) of both geometric reconstruction and semantic labeling.
Based on VSF, we select the next best views (NBV) as the target for each time step. 
We then jointly optimize the traverse path and camera trajectory between two adjacent NBVs, through maximizing the integral viewing score (information gain) along path and trajectory in the view space. Benefit from our online semantic reconstruction, our method achieves fast, accurate and complete scene parsing outperforming the state-of-the-arts. We have also conducted extensive experimental evaluations and comparisons to show that . 

To sum up, the contributions of this work are:
\vspace{-2pt}
\begin{itemize}
	\item A new approach to active scene understanding based on online semantic reconstruction.
	\item An efficient semantic segmentation network with incremental volumetric feature aggregation.
	\item A method for estimating the next best view based on the uncertainty in scene reconstruction and understanding.
	\item A method for joint optimization of robot path and camera trajectory in three-dimensional view space.
\end{itemize}

\section{Related Works}

\begin{figure*}[!t]
	\includegraphics[width=\linewidth]{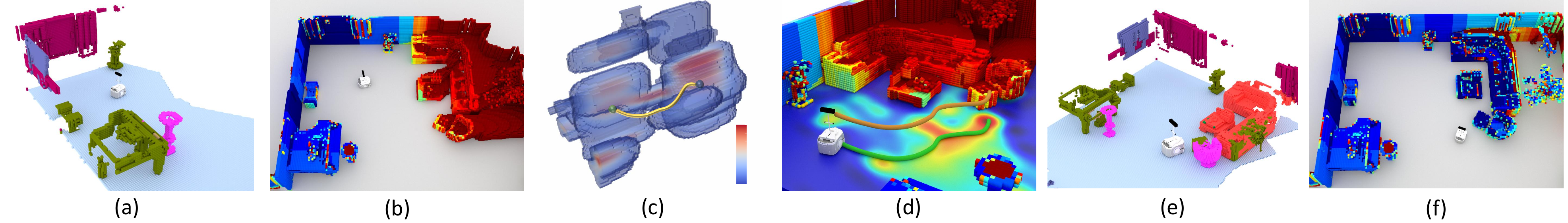}
	\caption{An overview of our method. Given the current reconstruction and understanding in (a), the robot performs online progressive reconstruction and entropy map computation/updating (b). Based on that, the view scoring field (VSF) is generated (c). Based on VSF, it performs field-guided optimization of robot path and camera trajectory (d). (e) shows the online reconstruction with semantic segmentation and (f) visualizes the updated entropy map for the next iteration.}
		
	\label{fig:overview}
\end{figure*}

\paragraph{Scene understanding}
Scene understanding has been a long-standing problem in both vision and graphics. The two main problems of scene understanding are scene classification and semantic parsing (object detection and/or segmentation). With the development of commodity depth sensors, the input of interest has been shifting from 2D RGB images~\cite{li2010object}, 3D CAD models~\cite{Fisher2011,Xu2014} or 3D point clouds~\cite{Nan2012}, to RGBD images~\cite{Gupta2013,song2016} and/or their 3D reconstruction~\cite{kim20133d,hou20183d,mccormac2017semanticfusion}. To take the advantage of deep learning, much attention has been paid on designing suited representation and efficient neural networks for the task of RGBD-based understanding~\cite{socher2012convolutional,Matterport3D,song2016ssc,qi20173d}. Most existing works have hereunto been devoted to offline, passive understanding based on the already acquired scene data. There are surprisingly limited works studying how to actively acquire scene data which are most useful for scene understanding, albeit the availability of real-time RGBD acquisition and reconstruction in nowadays.

\paragraph{Online RGB-D reconstruction}
With the introduction of commodity depth cameras, we have seen significant advances in online RGB-D reconstruction.
KinectFusion \cite{Newcombe2011,Izadi2011} was one of the first to realize a real-time volumetric fusion framework of~\cite{Curless96}.
In order to handle larger environments, spatial hierarchies \cite{chen2013scalable}, and hashing schemes \cite{Niessner2013,Kahler2015} have been proposed.
At scale, these methods also required robust, global pose optimizations which are common in offline approaches \cite{choi2015}; however, fast GPU optimization techniques \cite{Dai2017} or online re-localization methods \cite{Whelan2015} allow for real-time global pose alignment.
Our work builds upon this line of research to achieve active RGBD-based scene understanding.

\paragraph{Active object recognition}
Autonomous object detection and/or recognition is one of the most important ability of domestic robots.
A common solution to active object recognition is to actively resolve ambiguities of a certain viewpoint
in recognizing an object~\cite{zhao2018triangle}. 
In cases where the \emph{target object is known},
Browatzki et al.~\cite{browatzki2012active} define characteristic views, on a view sphere around the object, which are most beneficial in discriminating similar objects.
Potthast et al.~\cite{potthast2016active} introduce an information-theoretic framework combining two common techniques: online feature selection for reducing computational costs and view planning for resolving ambiguities and occlusions. Similar idea was also utilized in~\cite{Xu2016} for active, fine-grained object recognition.
Song et al.~\cite{Song2015} propose an information-theoretic approach based on 3D volumetric deep learning~\cite{Wu15}.
When \emph{target objects are unknown}, detection and recognition need to be solved simultaneously~\cite{liu2019recurrent}.
Ye et al.~\cite{ye2018active} propose navigation policy learning guided by active object detection and recognition.
The work in~\cite{liu2018} is the most similar in spirit to ours. They develop a data-driven solution to autonomous object detection and recognition with one navigation pass in an indoor room. 
The problem is formulated as an online scene segmentation with 3D models from a shape database serving as templates.
Our work frames the problem as online volumetric reconstruction and deep-learning-based voxel labeling.

\paragraph{Active scene segmentation}
Semantic segmentation of an indoor scene is critical towards accurate robot-environment interaction.
However, many existing approaches do not involve an online active view selection~\cite{yi2016automatic}.
Mishra et al.~\cite{mishra2009active} propose fixation-based active scene segmentation in which the agent segments
only one image region at a time, specifically the one containing the fixation point by an active observer.
Similar method is also studied in~\cite{bjorkman2010active} which integrates different cues in a temporal framework for improving object hypotheses over time.
Xu et al.~\cite{Xu15} present an autoscanning system for indoor scene reconstruction with object-level segmentation.
They adopt a \emph{proactive} approach where objects are detected and segmented
with the help of physical interaction (poking).
Yang et al.~\cite{yang2017view} study interactive indoor scene segmentation with automatic view selection.
In our system, scene segmentation is achieved by actively selecting the best view points and traverse paths
that maximally determine the volumetric labeling. 
\section{Method}
\setlength{\belowcaptionskip}{-2mm}

\subsection{Problem Statement and Overview}

\paragraph{Problem statement}
Given an indoor scene whose map is unknown, the objective of our system is to drive a ground robot mounted with an RGBD camera to explore and actively parse the scene into semantic objects. It is impossible to plan the complete scan path in advance since the map of the target scene is unavailable at the beginning, which makes it a chicken-and-egg problem. We therefore have to solve for scene understanding and path planning simultaneously. Existing approaches to active scene scanning usually take a ``scan and plan'' paradigm, which only take geometric but not semantic information into consideration when planning the robot scanning.
In this work, we frame the problem from online reconstruction with semantic segmentation and propose a novel ``scan, understand, and plan'' solution.

\paragraph{Method overview}
For the purpose of online scene understanding, we introduce a semantic segmentation network based on online volumetric reconstruction, inspired by~\cite{hou20183d}. The basic idea of our network is to first extract multi-view 2D features and then perform feature aggregation based on 3D convolution over the online reconstructed TSDF volume. Different from the offline scene understanding in~\cite{hou20183d}, the input for semantic labeling is dynamic due to the progressive scanning and online reconstruction. Therefore, the feature aggregation must follow the online reconstructed TSDF volume. Furthermore, to avoid redundant computation, our network bypasses the known and unchanged voxels in the TSDF volume during feature aggregation, thus significantly improving the online efficiency.

To guide the robot in achieving an fast online semantic reconstruction with minimal scanning effort, we adopt an information-theoretic approach to Next-Best-View (NBV) prediction through minimizing uncertainty (entropy) of semantic reconstruction. The entropy measures the uncertainty of both geometric reconstruction and semantic segmentation. In particular, we present a field-guided optimization of robot path and camera trajectory to maximize the information gain in traversing and scanning between every two adjacent NBVs.

An overview of the process is given in Algorithm~\ref{algo}. Any scanning move of the robot would collect some semantic information $S$ of the unknown scene. 
An entropy-based (section~\ref{Information_Entropy}) view scoring field $F$ is generated based on the online reconstructed TSDF with semantic labels $D$ (section~\ref{IncrementalSemantic}). To maximize the scanning efficiency, the Next-Best-View (NBV) should enable the robot to reduce the overall entropy as much as possible in the next move.
Based on the online updated entropy map and occupancy grid $T$, we compute a view scoring field $F$, based on which the robot path and camera orientation can be optimized jointly (section~\ref{path_planning}).
The above process repeats until the terminate condition is met. 

\begin{algorithm}[b!]\small
\caption{Robot scanning guided by online reconstruction and semantic segmentation.}
\label{algo}
\SetCommentSty{textsf}
\SetKwInOut{AlgoInput}{Input}
\SetKwInOut{AlgoOutput}{Output}
\Indm

\Indp
\AlgoInput{ Initial TSDF $D_0$, occupancy grid $T_0$ with few random scans and robot location $v_r$ }
\AlgoOutput{ Semantic label $S$ and optimized scanning path $\{P_i,C_i\|i=0...k\}$}
Initialize $S_0 \leftarrow f_{rec}(D_0)$\;
Initialize entropy map $H_0$ from $T_0$ and $S_0$
\;
Initialize view scoring field $F_0 \leftarrow H_0, T_0$
\;
\Repeat {Terminate condition is met} {
    \tcp{Path planning and camera rotation optimization based on $F$}
    Find NBV $v_i \leftarrow \arg\max_{v}F_i$\;
    Find the optimal robot and camera path $P_i, C_i$ from $v_r$ to $v_i$\;
    \tcp{Update scene mapping based on given path}
    Scan along $P_i, C_i$ and update semantic map $S$\;
    Update $S_i\leftarrow f_{rec}(D_i)$\;
    Update $H_i$ from $T_i$ and $S_i$\;
    Update $F_i$ from $H_i, T_i, H_{i-1}, T_{i-1}$\;
    \tcp{Record current path planning}
    $\{P_i,C_i\|i=0...k\} \leftarrow P_i,C_i$\;
}
\Return $S$, $\{P_i,C_i\|i=0...k\}$\;
\Indm
\end{algorithm}
\DecMargin{0.5em}

\begin{figure*}[t]
		\includegraphics[width=\linewidth]{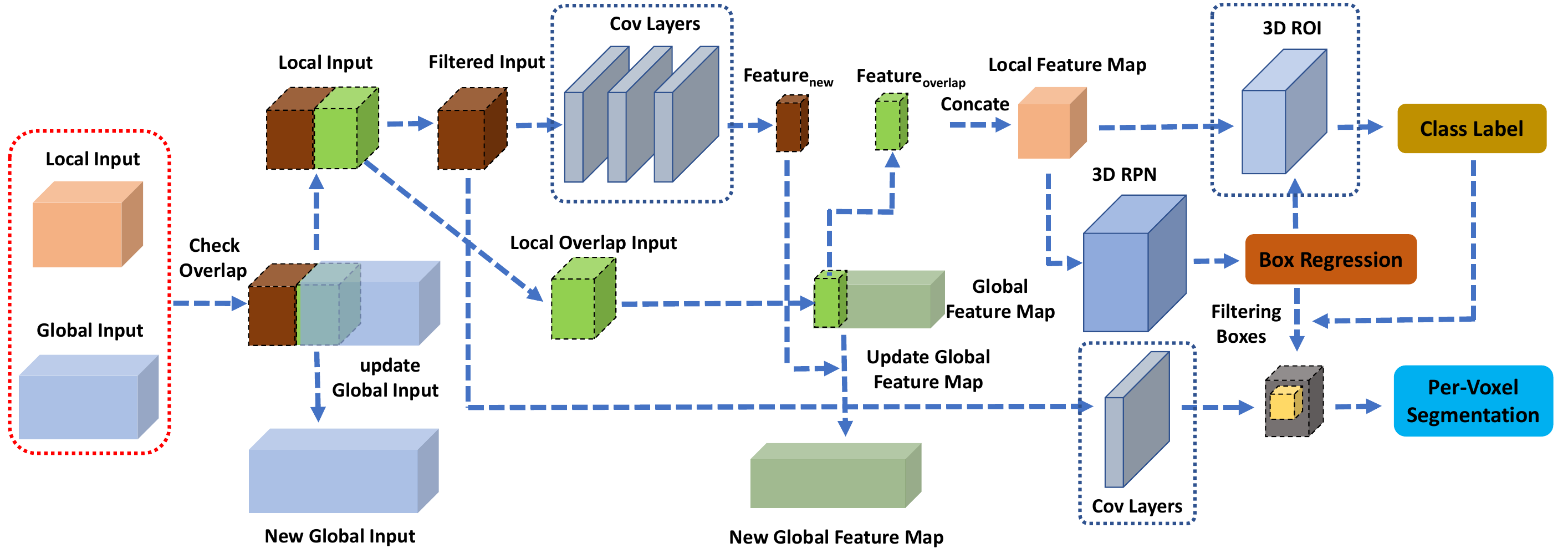}
	\caption{The architecture of our online semantic segmentation network. Note that the key difference between our network and 3D-SIS\cite{hou20183d} is that our feature aggregation is incremental which fits the online processing more natural. 
	The input of our network contains two components, which are local and global (red box). We save massive processing time on redundant operations of overlapped areas. More specifically, the voxels which have been observed in previous steps would not go through the 3D convolution layers and 3D-ROI (blue box) again. Our network would re-unitize the stored information directly. 
	}
	\label{fig:network}
\end{figure*}

\subsection{Online Reconstruction with Semantic Segmentation}
\label{IncrementalSemantic}
We measure the quality of a scan view 
by how much the uncertainty of scene understanding would be reduced through this move. In our work, the uncertainty of scene understanding is measured from two aspects, i.e., geometry reconstruction and semantic segmentation.

\paragraph{RGBD-based reconstruction with volumetric representation} Given a sequence of RGBD images, we adopt the volumetric representation (TSDF) for depth fusion~\cite{Curless96}. The construction of TSDF $D$ is incremental. The occupancy uncertainty of each voxel $v$ is reduced when more images are fused into $D$. Usually, the occupancy of $v$ can be modeled based on a 1D half normal distribution: $t(v) = 1-\left |  X \right |, X \sim N(0,\sigma^2(v))$. 
The variance $\sigma^2(v)$ provides a measure of reconstruction uncertainty. More specifically, the variance $\sigma(v)^2$ is defined based on how many images provide positive support for the occupancy of $v$~\cite{Hornung13}. The positive support here means the camera gets a reflected signal from $v$ when shot a depth image and vice versa. 
To make it simple, every positive support $i$ would provide $\sigma_{occ}(v,i)=0.85$ and every negative support $i$ would provide penalty $\sigma_{free}(v,i)=-0.4$.

\begin{equation}
\label{geometryUncertainty}
    \sigma(v) = \sum_i\sigma_{occ}(v,i) + \sum_i\sigma_{free}(v,i)
\end{equation}

\paragraph{Semantic reconstruction network}
To incrementally gain semantic information during scanning, we propose a network to predict a 3D semantic segmentation based on the TSDF $D$. More specifically, we want to infer the semantic labeling over the TSDF on a per-voxel basis. The backbone of our network is similar to~\cite{hou20183d}. We first briefly review the network architecture and then discuss our improvement over it.

The network is composed of two main components including object detection and per-voxel labeling prediction. 
Each of these component has its own feature extraction module. Each module is composed of a 2D and 3D feature extraction layers. The extracted 2D and 3D features are aggregated by a series of 3D convolutional layers over the TSDF volume. The object detection component comprises a 3D region proposal network (3D-RPN) to predict bounding box locations, and a 3D-region of interest (3D-RoI) pooling layer followed for classification. The per-voxel mask prediction network takes geometry as well as the predicted bounding box location as input. The cropped feature channels are used to create a mask prediction for per-voxel semantic labeling as well as the confidence score.

However, this network is designed for offline scene understanding where the reconstruction is already given. In our problem setting, the online reconstruction is executed online, with smooth and progressive RGBD acquisition. This means that there is immense overlap ($>30\%$) between the observations of every two adjacent RGBD frames. Directly applying feature aggregation would result in much computational redundancy. To support real-time application, we make a modification to this network to reuse the previous feature aggregation as much as possible.

\begin{figure}[t]
	\includegraphics[width=\linewidth]{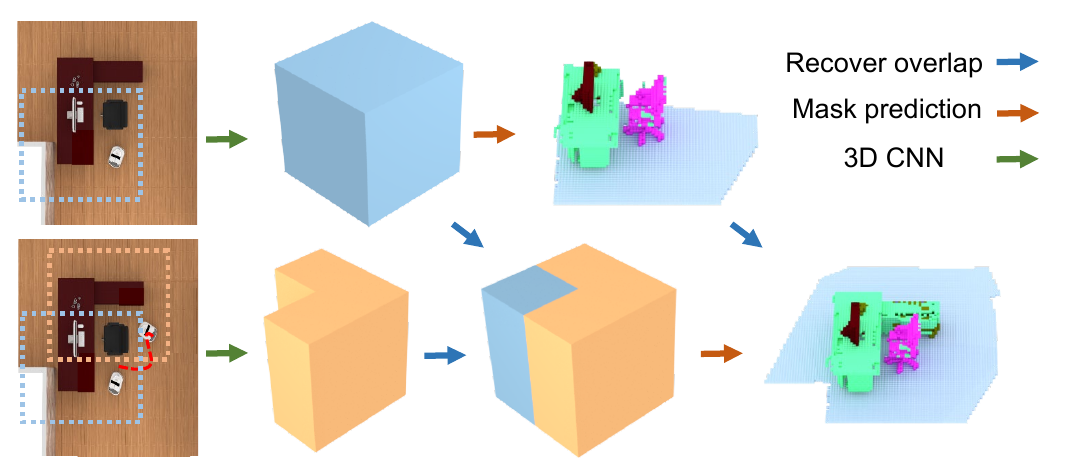}
	\caption{An illustration of incremental semantic segmentation.}
	\label{fig:incremental}
\end{figure}

\paragraph{Incremental 3D feature aggregation} Most offline scene understanding methods do not consider how to process dynamic inputs, we present an incremental semantic segmentation network specifically designed for online understanding. The key insight of our approach is that 3D convolution should be performed only on the newly-observed voxels and reuse the previous result for overlapping areas as much as possible. Figure~\ref{fig:incremental} gives an illustration of this.

More specifically, we maintain a global data structure to record the TSDF and 3D features information for all the observed voxels. When the network gets a new local input, the first step is finding the overlapped areas between the input and the global record. And our proposed network would skip the 3D convolution and reuse the stored information directly for this overlapped areas which can save a lot of computational time.

Moreover, our network would also reuse the results of the 3D-RPN. All the box proposals in the overlapped areas would not be processed again for a new local input. Note that we only reuse the features of those voxels whose convolution is performed completely inside the overlapping volume. For the boundary voxels where the feature convolution involves both old and incremental voxels, we do not reuse but instead recompute their features. Even by doing this, our method still saves a lot of computation since boundary voxels represent only a small portion of the overlapping volume. By removing these redundant proposals, our incremental network would improve the efficiency one step further. In our experiment, the incremental process would make our network be 23.6\% faster if the input has 50\% overlapped area and 41.1\% faster if the input has 75\% overlapped area. More details about out network can be found in Figure~\ref{fig:network}.



\subsection{Reconstruction and Segmentation Entropy}
\label{Information_Entropy}

We adopt Shannon entropy to measure the information gain of robot scanning. In particular, we estimate the average new information the robot can collect under a specific pose. In other words, we want to measure how much uncertainty would be reduced by a potential scanning view. 

The entropy map $H$ is defined on each voxel in the 3D scene. Different from previous method like ~\cite{bai2016information}, we do not only count the geometry occupancy possibility of each voxel $v$ but also the predicted semantic label as new information. The general definition of entropy in our problem is $H(v)= -\sum p(v) \log p(v)$ and we can measure the gained information $I$ as $I(v)=H(v)-H(v|v^{\text{new}})$. The key point to evaluate the quality of new information through $I$ is how to define probability $p$ in $H$ for geometry and semantic information respectively. Then we can sum these two item up in a weighted fashion to get final formulation of the gained information. $\alpha$ and $\beta$ are constants to weight the geometry term and the semantic term.
\begin{equation}
\label{gainedInfo}
     I(v)=\alpha I_{semantic}(v) + \beta I_{geometry}(v)
\end{equation}

\paragraph{Geometry reconstruction entropy} As introduced in Section~\ref{IncrementalSemantic}, the uncertainty of voxel $v$ in geometry reconstruction can be defined as Equation (\ref{geometryUncertainty}). However, the output range of this uncertainty formulation is $\left [ 0,\infty  \right ]$ which can not be adopted as the probability function $p$ in a entropy formulation directly. We simply map this uncertainty function to $\left [ 0,1  \right ]$ as below and use it in our geometry reconstruction entropy.
\begin{align}
     I_{geometry}(v) = H_{g}(v)-H_{g}(v|v^{\text{new}})\\
     H_{g}(v)=-p_g(v) \log p_g(v)-(1-p_g(v)) \log (1-p_g(v)),\\ p_g(v)=\frac{e^{\sigma^2(v)}}{1+e^{\sigma^2(v)}}
\end{align}

\begin{figure}[!t]
	\includegraphics[width=\linewidth]{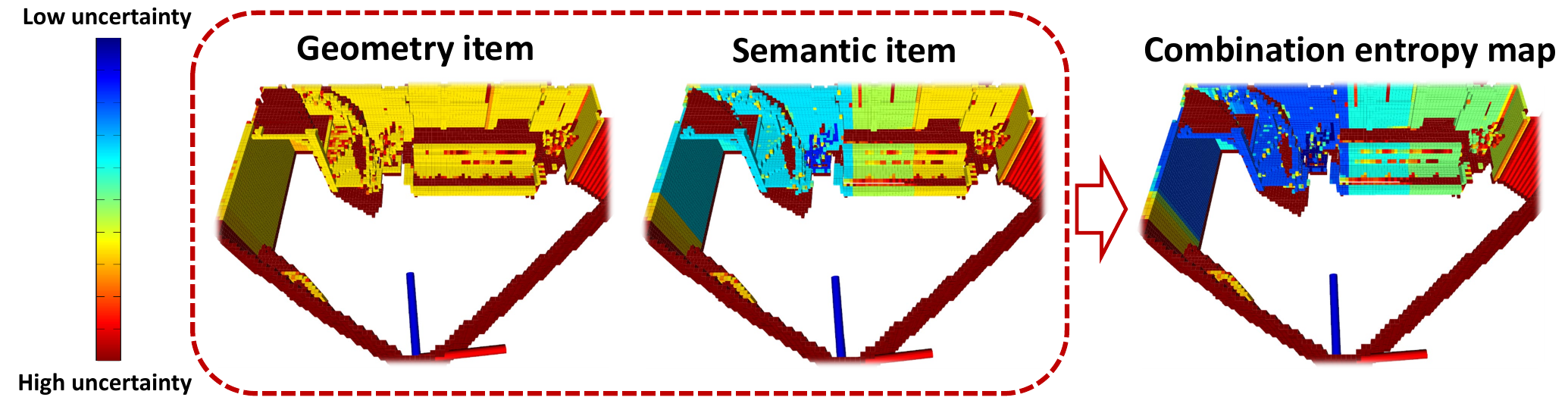}
	\caption{Visualization of scanning entropy encompassing both geometric and semantic uncertainty.}
	\label{fig:visual-entropy}
\end{figure}

\paragraph{Semantic segmentation entropy} To measure the uncertainty of semantic segmentation, we would like to consider both the predicted semantic label and corresponded confidence score for a voxel $v$ into consideration. If the predicted semantic label for $v$ in current scan move keeps the same as the previous predicted result and the confidence score becomes higher, then the uncertainty for semantic segmentation for $v$ is reduced. 
In another case, we can gain more information from $v$ if the confidence score is higher even the predicted semantic labels are different. Therefore, we have a following formulation for semantic segmentation entropy, where $p_s$ represents score of semantic prediction given by our semantic reconstruction network for a specific label $s$.

\begin{equation}
\scalebox{0.8}{$I_{semantic}(v)=\begin{cases}
&\sum_s p_s(v|v_{\text{new}}) \log p_s(v|v_{\text{new}}) - \sum_s p_s(v) \log p_s(v), \quad \text{if } S_{v}=S_{v^{\text{new}}}\\
&-\sum_s p_s(v|v^{\text{new}}) \log p_s(v|v^{\text{new}}), \quad \text{if } S_{v}\neq S_{v^{\text{new}}} \text{ and }p_{v}<p_{v^{\text{new}}}\\
&0, \quad \text{otherwise}
\end{cases}$}
\end{equation}
where $v^{\text{new}}$ denotes the new observation for voxel $v$ and $S_v$ is semantic label of $v$.

In Figure~\ref{fig:visual-entropy}, we illustrate this idea of combined entropy.
Here, higher entropy value means lower confidence since more information is required for more reliable predictions. 
This becomes clear when we only consider the geometry term; in this case, the robot cannot predict a meaningful next step since the uncertainty for the the geometry reconstruction is similar everywhere.
The semantic term resolves this ambiguity, and as such it provides good guidance about which area the robot should focus on in the next move.
In the concrete case of Figure~\ref{fig:visual-entropy}, there is a valid semantic object (sofa) in the right view.

In order to highlight the benefits of our combined entropy in the context of semantic labeling quality, we provide a side-by-side
visual comparison in Figure~\ref{fig:example-entropy}. 
In this way, the objective of our NBV prediction is clear now. 
The NBV should be the view that all the voxels in it have the highest uncertainty. 
Based on Equation (\ref{gainedInfo}), we have a formulation for NBV prediction as following where $\Omega(v,n)$ represents all the voxels in current camera view $n$:
\begin{equation}
\label{oldnbv}
NBV = \arg\max_{n} \sum_{k\in\Omega(v,n)} \alpha I_{semantic}(k) + \beta I_{geometry}(k)
\end{equation}




\subsection{Field-guided Scan Planning}
\label{path_planning}

\begin{figure}[t]
	\includegraphics[width=\linewidth]{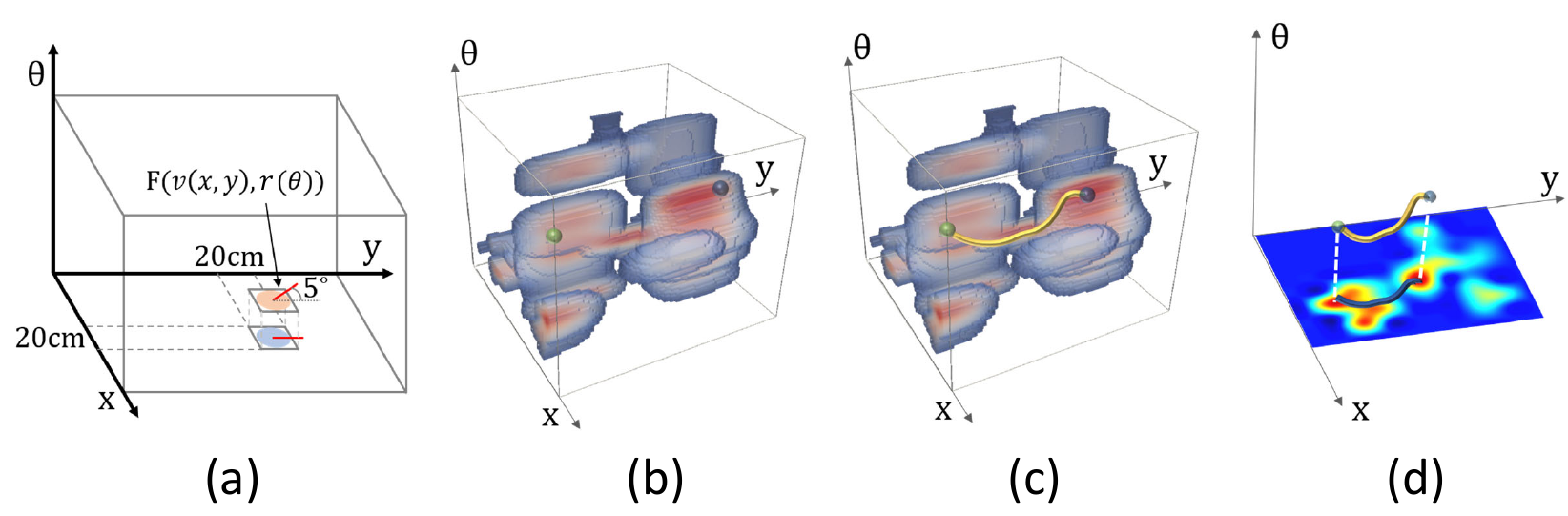}
	\caption{(a): Illustration of 3D parametric space of location and orientation.
             (b,c): Visualization of view scoring field and the optimal path found by $A^*$ algorithm between two camera poses in the field.
             (d): The computed robot trajectory based on field-guided optimization.}
	\label{fig:vsf}
\end{figure}


\paragraph{View scoring field} Our scan planning is composed of two components, NBV prediction and path planing with camera optimization. These two components are implemented upon a 3D field which records the entropy information described in section~\ref{Information_Entropy}. Please note that this field is incrementally constructed with the scanning process.

Besides information gain, the field construction accounts for the following factors:

\begin{itemize}[itemindent = 15pt]
    \item Safety: View point must be in free space and keep a safe distance away from obstacles;
    \item Visibility: Views should orient toward objects or frontiers to maximize information gain;
    \item Movement cost: Robot traverse path should be as short as possible.
\end{itemize}

In this case, we find Equation (\ref{oldnbv}) is not sufficient to find the most appropriate NBV for our system.
In addition to the gained information $I$, we introduce the occupancy grid $T$ to measure the value in our view scoring field which would be helpful to measure the above three factors.


To ensure robot safety, we get obstacle information from the 2D projection of $T$, and only sample views which can keep a safe distance $0.35m$ from obstacles. 

We further consider the visibility to frontier. Frontier is the boundary between (known) empty regions and unknown ones, which is a well known driving factor for robot exploration in robotics. 
To this end, we measure the visibility to frontier by counting the frontier voxels visible in the current view frustum. Specifically, it is estimated based on $T$:

\begin{figure}[t]
	\includegraphics[width=\linewidth]{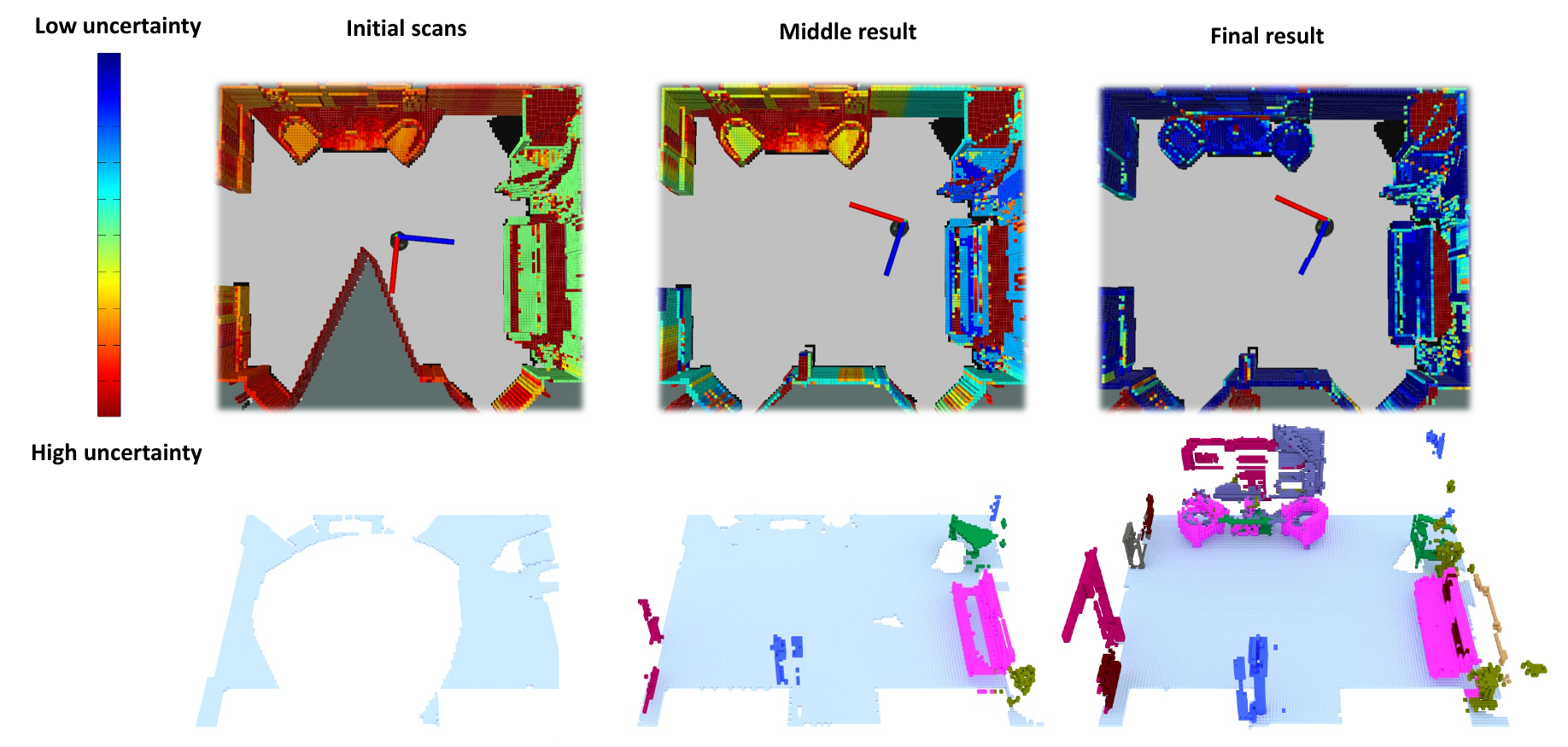}
	\caption{The evolution of scanning entropy over increasing scans.}
	\label{fig:example-entropy}
\end{figure}

\begin{equation}
V(v,r)=\sum_{k\in\Omega(v,r)}T(k),
\quad
T(v)=
\begin{cases}
1, &\quad \text{if voxel }v \text{ is frontier} \\
0, &\quad \text{otherwise}
\end{cases}
\end{equation}


where $r$ is the given view 
from voxel $v$ and $\Omega(v,r)$ means all the voxels in this view frustum. 
We need to ensure that the planned path is not too long. 

Here, we use an approximate distance estimation formulation in order to enforce this movement constraint. $L(v) = e^{-\frac{\text{dist}^2(v,v_{robot})}{2\sigma^2}}$ 
, where $v_{robot}$ is the current robot location. After formulating all these factors, we will discuss the details about how to assemble them to get 3D view scoring field. For each ground grid voxel $v$ of the given scene, we sample some different views. And the safety, visibility and movement factor are calculated for each view $r_i,i\in\{0...k\}$ of every $v$. And we will have the final view score formulation for each grid voxel $v$ with different camera view $r_i$, and we have a 3D visualization of this field in Figure~\ref{fig:vsf}(a):

\begin{equation}
\label{eq:nbv}
F(v, r_k)=\frac{\alpha V(v,r_i)+\sum_{v\in\Omega(v,r_i)}I(k)}{L(v)}
\end{equation}

\paragraph{Optimization formulation} We will update this view scoring filed $F$ after each scan move, and the NBV can simply be computed by the optimization $NBV = \arg\max_{v,r} F(v,r)$. The main challenge in this part lies in how to compute a collision-free path from the current robot position $v_{robot}$ to the $NBV$, such that the path maximizes the information gain of semantic reconstruction and minimizes the traverse distance.

In order to guarantee robot safety and scanning efficiency, the view scoring filed $F$ plays a significant role in path planning algorithm. 
Formally, we define $C(P)$ as the total cost of the optimal path $P$:
\begin{equation}
\label{pathpEq}
C(P)=\inf_{\pi\in\Pi (v_{robot},v_i)}\int \eta-F(v,r)d\pi(v,r)
\end{equation}
\begin{equation*}
\text{s.t.} \quad V_{\text{rotation}}>\frac{\delta \pi}{\delta c}
\end{equation*}
where $\Pi(v_r,v_i)$ is the set of all possible paths from location $v_r$ to $v_i$, $V_{\text{rotation}}$ is maximum rotation speed of the robot camera and $\eta$ is a big constant, which we set to 500 in our experiment. 
which helps to adopt $F$ in our costmap. However, even we consider the safety factor when designing the view scoring field $F$, it is still can not guarantee that the found path through Equation (\ref{pathpEq}) is collision-free. we introduce a 2D obstacle costmap to enhance $F$ to solve this problem.

\begin{figure*}[t]
	\includegraphics[width=\linewidth]{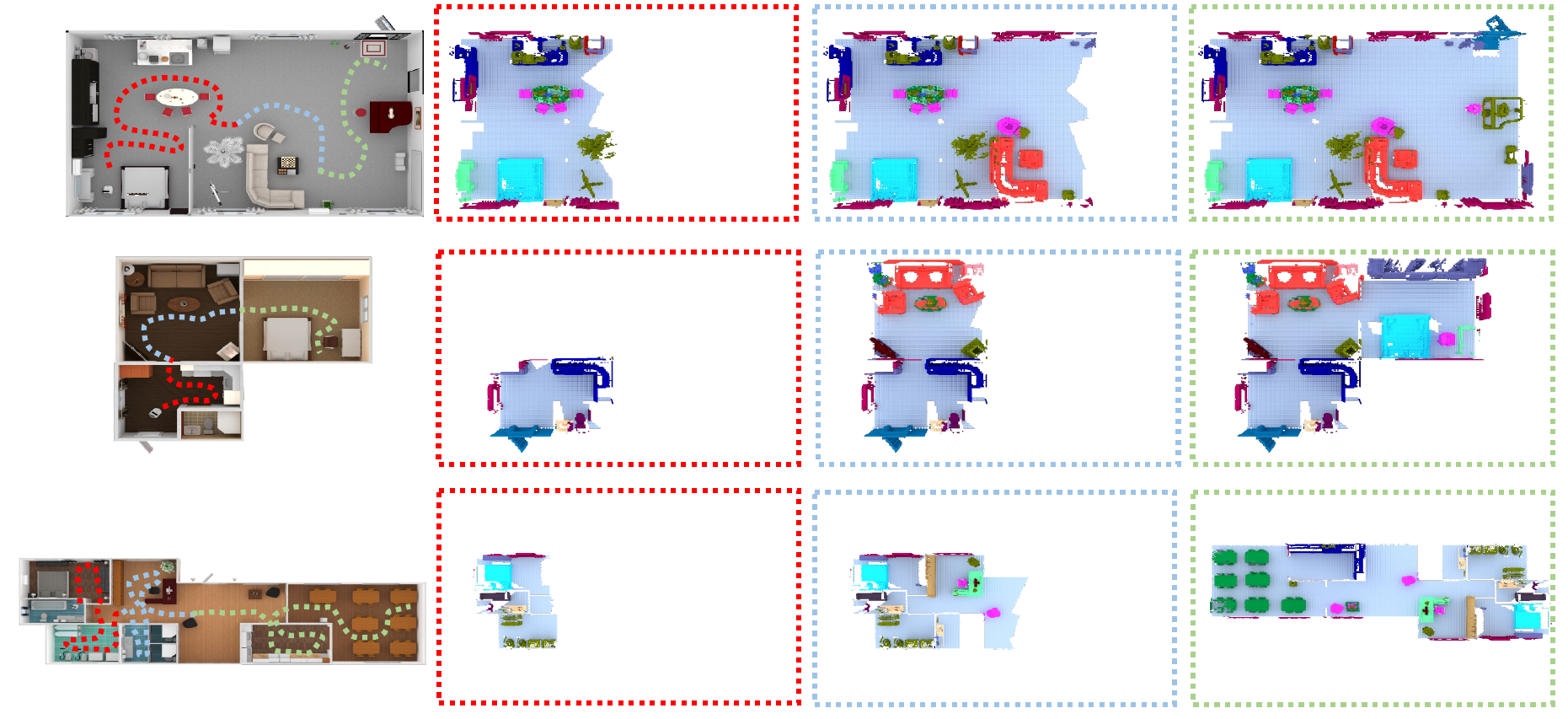}
	\caption{Visualization of our active scene understanding process for three different scenes.}
	\label{fig:scanProcess}
\end{figure*}

The 2D obstacle map $f_o$ is obtained from projection of 3d occupancy grid map $T$. We generate the obstacle cost map by using two-dimensional Gaussian distribution.

\begin{equation}
f_o(v) = e^{-\frac{{\min_{v_k\in\{v\|T(v)=1\}}}\text{dist}^2(v,v_k)}{2\sigma^2}}
\end{equation}



We integrate the 2D obstacle map $f_o$ into the view scoring field $F$ and change the optimization formulation from Equation (\ref{pathpEq}) to the following:
\begin{equation}
\label{pathpEqnew}
C(P)=\inf_{\pi\in\Pi (v_{robot},v_i)}\int \eta-F(v,r)d\pi(v,r)+\int \eta f_o(v,r)dv
\end{equation}
\begin{equation*}
\text{s.t.} \quad V_{\text{rotation}}>\frac{\delta \pi}{\delta c}
\end{equation*}

\paragraph{Scan planning by optimization}
To solve the path and camera optimization defined in Equation (\ref{pathpEqnew}) 
, we adopt $A^*$ algorithm to find the optimal solution in discrete level. Figure~\ref{fig:vsf} illustrates how we get the camera view path and the robot path from the optimal path given by the 3D costmap. We project the optimal path which given by the $A^*$ algorithm to $\theta$ axis to get the camera rotation sequence and the projected path on $xy$ plane is the optimal 2D robot path.

This ``scan, analyze, and plan'' process is repeated until the terminate
condition is met, leading to a progressive understanding by the robot. In our experiment, the robot will stop the exploration if the overall entropy $\sum_v I(v)$ is reduced below a certain threshold, which is set to 150 in our experiment.



\section{Results and Evaluation}
\label{sec:eval}

There are three primary questions that we seek to answer with our experiments and evaluations.

\begin{itemize}[itemindent = 15pt]
	\item How does our approach compare to previous work in terms of distance traveled, time cost, and semantic prediction quality?
	\item How much effect does the semantic entropy item have over the results?
	\item How much does field guided path planning improve the scanning efficiency?
\end{itemize}

\subsection{System and implementation}
\paragraph{Simulation setup}
The simulation is conducted by using the Gazebo simulator~\cite{gazebo}. We adopt a differential drive ground robot equipped with a virtual RGB-D camera simulating the Kinect v1 sensor. We assume the sensor has a depth range of $[0.5, 4.5]$m with Gaussian noise, which $\mu = 0$ and $\sigma = 0.03$. 
The camera is mounted on top of the robot and has one DoF of azimuth rotation.
In order to achieve a realistic simulation, the ground robot will obtain a noisy pose estimation from the simulator.
The simulation runs on a computer with an Intel I7-5930K CPU (3.5GHZ *12), 32GB RAM, and an NVIDIA GeForce GTX 1080 Graphics card.

\begin{table}[t]
\caption{NBV estimation time comparison between our method and state-of-art methods, NBO~\cite{liu2018}, BO~\cite{bai2016information} and IG~\cite{Charrow2015}.}
\centering
\scalebox{1.0}{
\begin{tabular}{ccccc}
\hline
\multicolumn{1}{|c|}{Method}                                                        & \multicolumn{1}{l|}{Ours} & \multicolumn{1}{l|}{NBO}  & \multicolumn{1}{c|}{BO}   & \multicolumn{1}{c|}{IG}   \\ \hline
\multicolumn{1}{|c|}{\begin{tabular}[c]{@{}c@{}}Cost time\end{tabular}} & \multicolumn{1}{l|}{5.5s} & \multicolumn{1}{l|}{7.4s} & \multicolumn{1}{c|}{6.6s} & \multicolumn{1}{c|}{4.3s} \\ \hline
\end{tabular}}
\label{tabNBV}
\end{table}

\paragraph{Dataset}
Our benchmark dataset is built upon the virtual scene dataset SUNCG. SUNCG contains 40K human-modeled 3D indoor scenes with visually realistic geometry and texture. It encompasses indoor rooms ranging from single-room studios to multi-floor houses. We select 180 scenes which are suitable for navigation and exploration task. These scenes have on average $4.5$ rooms and the average room area is $45$m$^2$. Different interiors including offices, bedrooms, sitting rooms, kitchens, etc. are involved in our dataset to guarantee the test variety. The dataset also provides ground truth object segmentation and labeling for the scenes.

\begin{figure}[t]
\includegraphics[width=\linewidth]{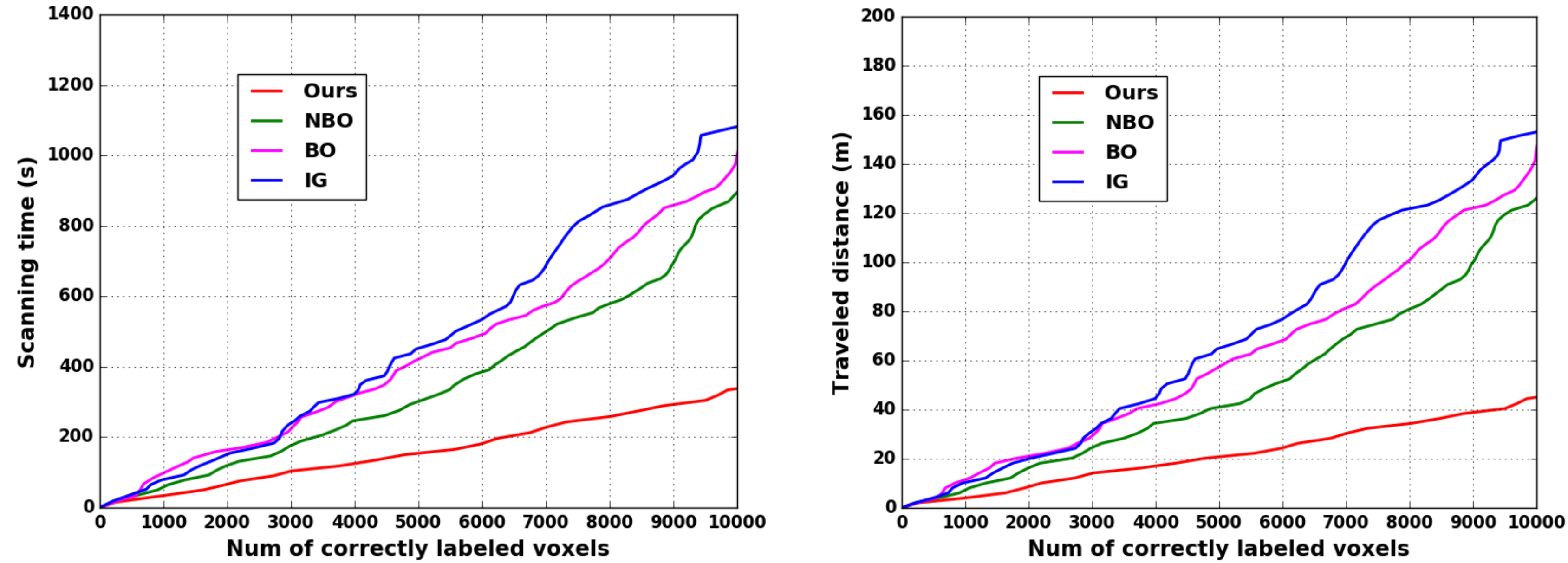}
\caption{Comparing scanning efficiency between our method (red), NBO (green), BO (magenta), and IG (blue) in different scenes. It is measured by traveled distance and time over numbers of correctly labeled voxels.}
\label{fig:nbvComp}
\end{figure}

\begin{figure}[t]
	\includegraphics[width=\linewidth]{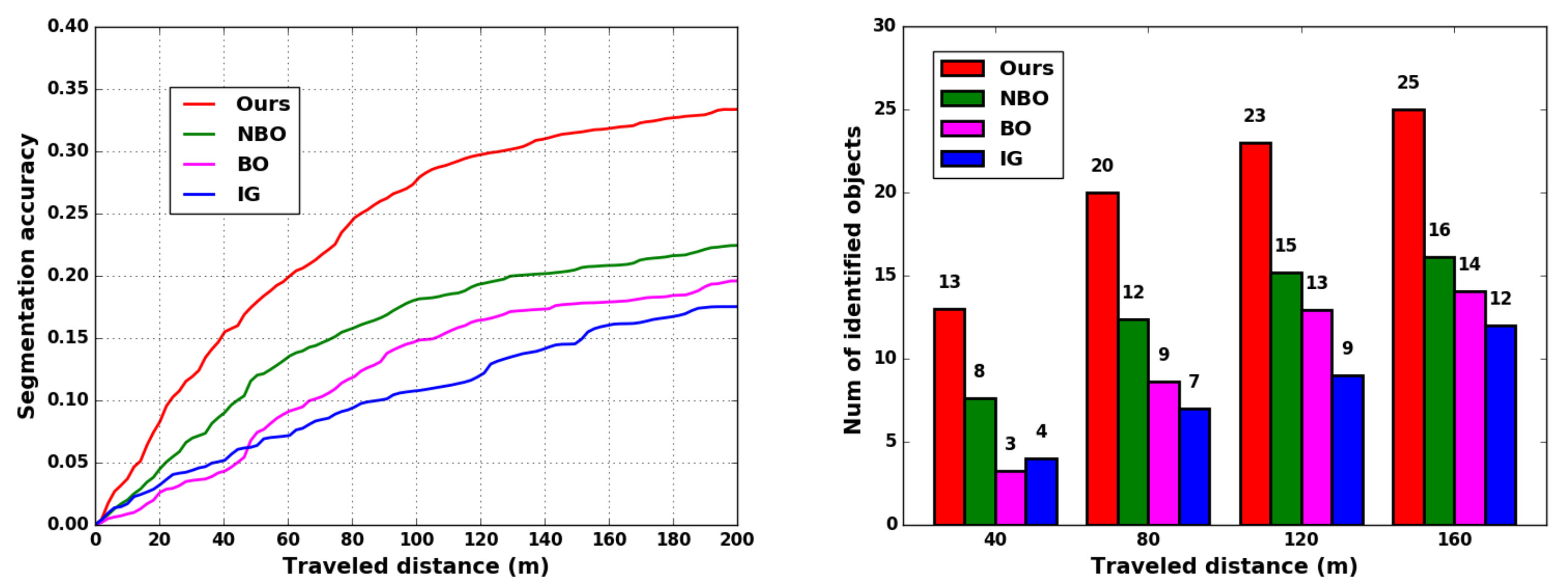}
	\caption{Comparing segmentation accuracy (left) and object recognition (right) against NBO, BO, and IG. Note that the numbers of total semantic voxels and objects can be obtained from ground truth labels.}
	\label{fig:nbvCompAcc}
\end{figure}

\paragraph{Parameters and details}
The 3D occupancy grid $T$ is constructed with a resolution of $0.05$m. The resolution of viewing score field is $0.4$m. In our experiments, the upper limit of linear and angular speed of the robot is $0.3$m/s and $40^\circ$ per second, respectively. The coefficient ratio $\alpha$ is set to $1.0$ and $\beta$ is set to $0.3$.

\subsection{Comparison and evaluation}
In this section, we conduct a series of experiments and comparisons which focus on evaluating scanning efficiency and semantic mapping quality of our method. Since it is impossible to get the input scene fully labeled in voxel-wise, we evaluate the scanning efficiency by measuring the time for our system to achieve a given mass of correctly labeled voxels. To evaluate semantic quality, we measure the accuracy of final scene segmentation.

\paragraph{Comparison with alternative NBV methods}
Our method is compared to several state-of-the-art NBV techniques: Bayesian optimization-based exploration method (BO)~\cite{bai2016information}, information-theoretic planning approach (IG)~\cite{Charrow2015} and Object-Aware based scene reconstruction algorithm (NBO)~\cite{liu2018}. For all these methods, a fixed forward-looking virtual camera is used.

\paragraph{Scanning efficiency}
We compare the scanning time and travel distance from the four kinds of approaches, while scanning the scenes virtually. The initial positions and orientations of the robot in all these methods are the same. The comparison about scanning time and traveled distance over correctly labeled voxel number are plotted in Figure~\ref{fig:nbvComp}. We observe that the scanning cost time and traveled distance are increasing as the scene semantic mapping gets more complete (more and more occupancy voxels get labeled). But the proposed approach always gets the least time and shortest distance.

To further discuss scanning efficiency of different NBV methods, we also compare the NBV estimation time (average cost time for NBV computation) in Table~\ref{tabNBV}, which is averaged over all 180 scenes we have tested. We found that, the most efficient method is~\cite{Charrow2015} due to the low complexity of its algorithm. However, its segmentation performance is the worst. \cite{liu2018} is built upon semantic prediction as our proposed method but it takes the most computation time and its performance is not as good as our method.

\paragraph{Semantic segmentation performance}
To evaluate the quality of semantic segmentation, we measure the segmentation accuracy and identified objects number (exclude wall, ceiling and floor) respectively. 
The segmentation accuracy and identified objects number over traveled distance is plotted in Figure~\ref{fig:nbvCompAcc}. 
The number of correctly labeled voxels are increasing while the robot explores more area, and the accuracy increasing as well. 
From these results, we can clearly see that our method achieves the highest semantic accuracy and maximum number of identified objects almost all the way.

\begin{figure}[t]
	\includegraphics[width=\linewidth]{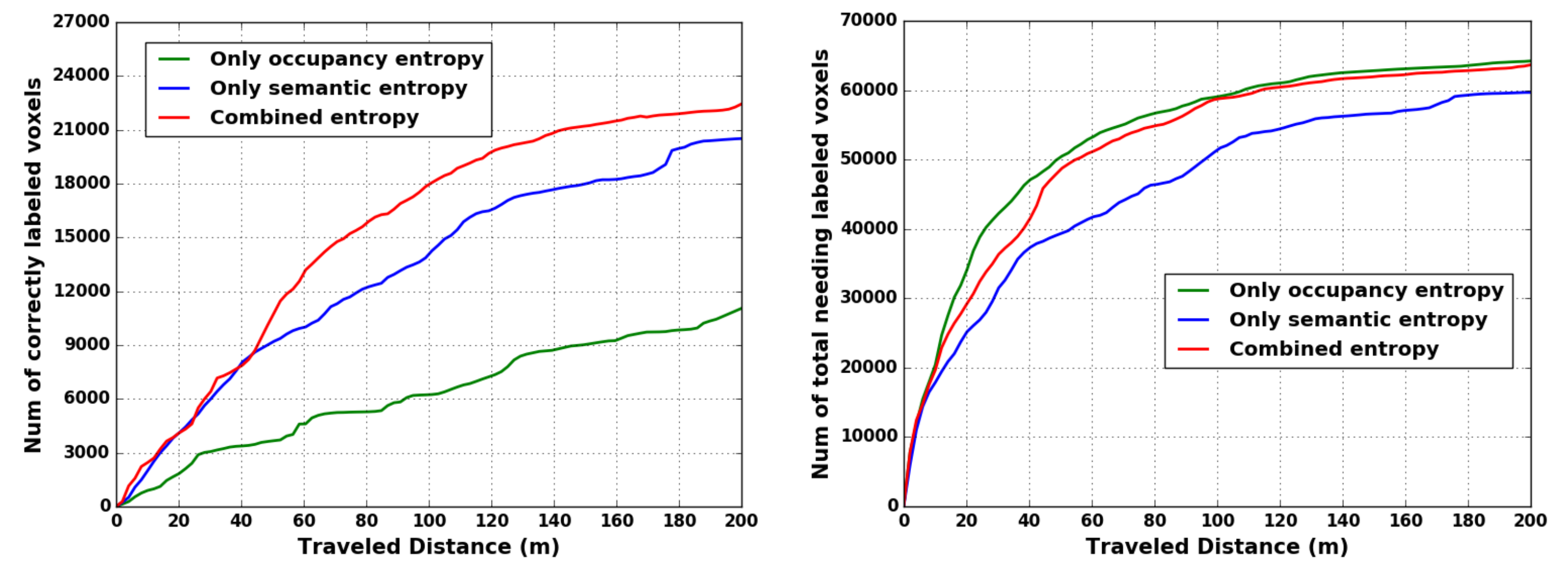}
	\caption{Effect of various entropy items on semantic segmentation performance (left) and exploration efficiency (right). The combined entropy leads to faster scene exploration and more complete segmentation results.}
	\label{fig:compareEntropy}
\end{figure}

\begin{figure}[t]
	\includegraphics[width=\linewidth]{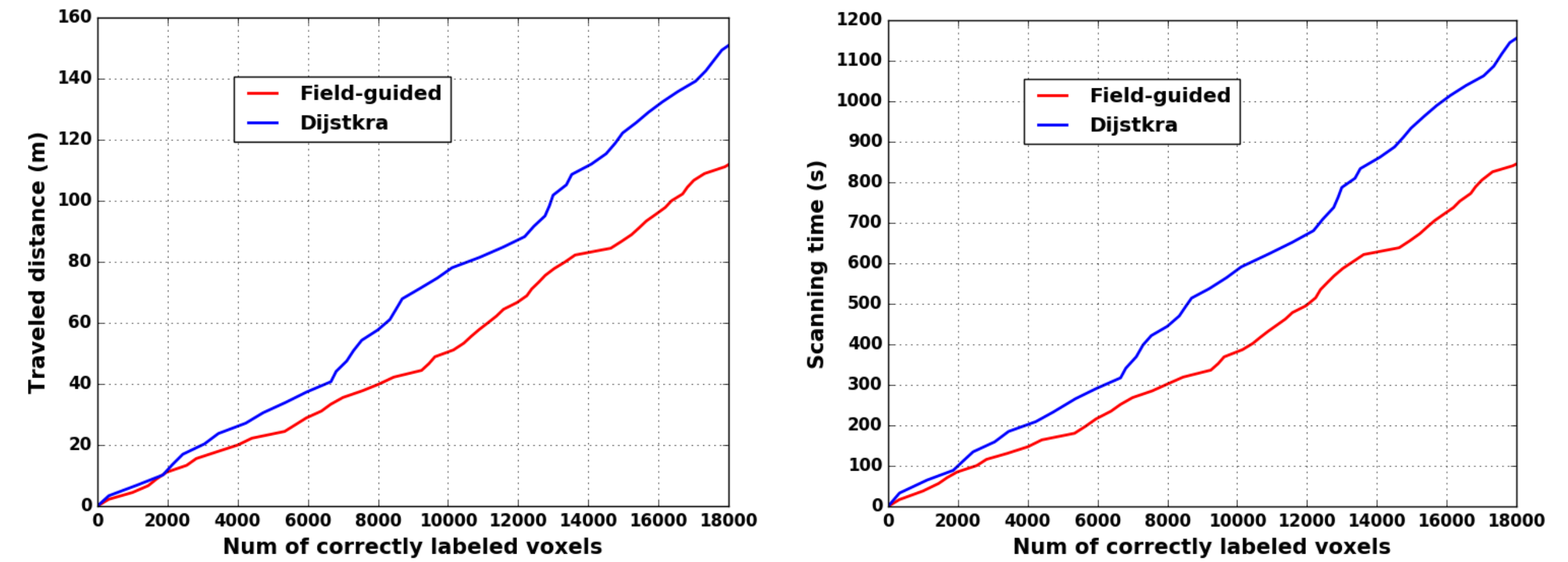}
	\caption{Effect of field-guided path planning on scanning efficiency. The proposed algorithm (red) is compared against classical Dijstra (blue) method. Using field-guided path planning, the robot travels less distance and time.}
	\label{fig:comaprePath}
\end{figure}

\begin{figure*}[t]
	\includegraphics[width=\linewidth]{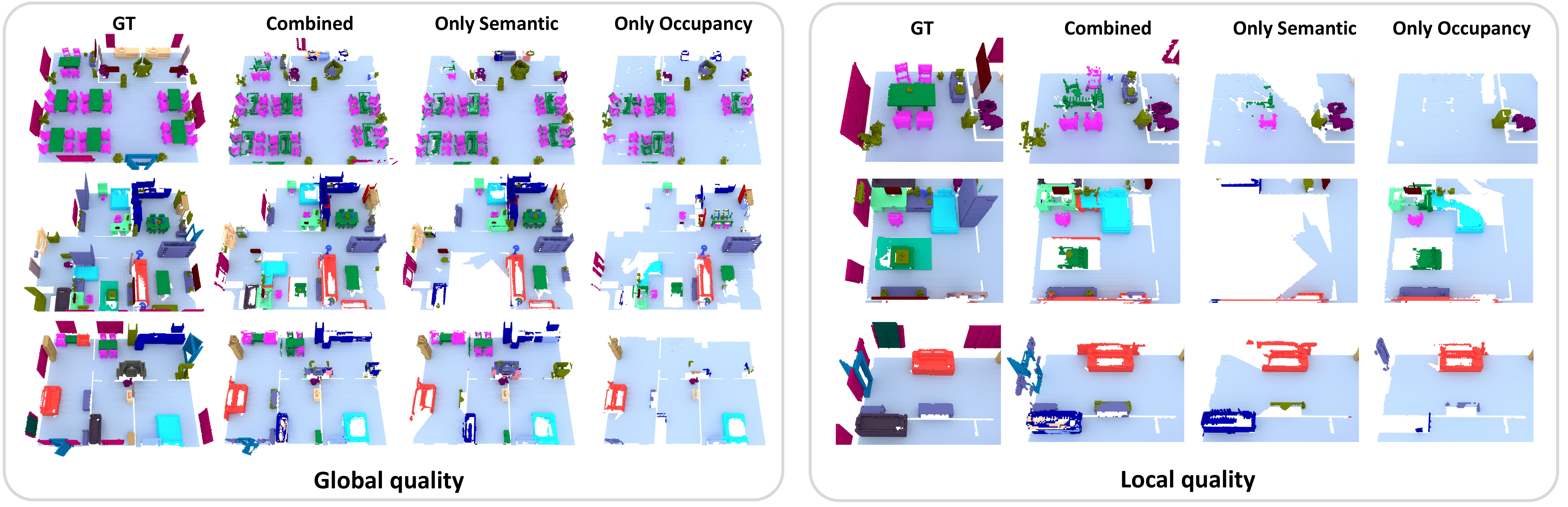}
	\caption{Qualitative comparison of semantic segmentation results with different entropy items (Left: global quality; Right: local quality).}
	\label{fig:entropyQuality}
\end{figure*}

\begin{figure*}[t]
	\includegraphics[width=\linewidth]{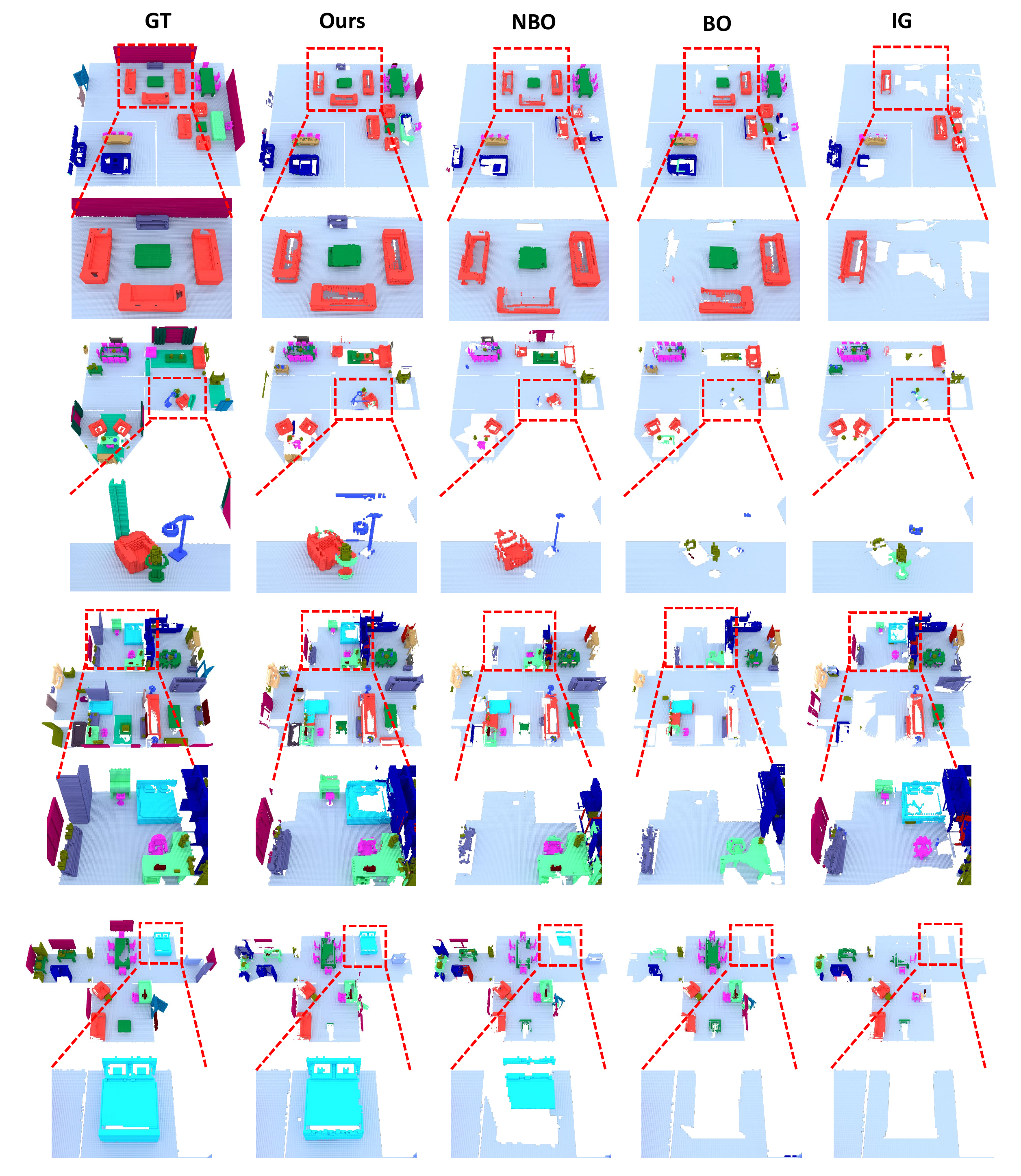}
	\caption{Qualitative comparison of indoor scene semantic segmentation on SUNCG dataset. Note that different colors represent different semantic labels.}
	\label{fig:nbvQuality}
\end{figure*}

To demonstrate the benefits of our algorithm one step further, we show visual results of the final semantic segmentation in Figure~\ref{fig:nbvQuality}. 
The results show that our scanning strategy leads to more complete and better results. For more visual results, please refer to the supplemental material.

\paragraph{Ablation study on semantic entropy}
Occupancy entropy tends to guide robot to explore more unknown space, while semantic entropy is more likely to guide robot to exploit scanned region. In this experiment, we investigate the effect of the semantic entropy item on the semantic segmentation efficiency and quality. Figure~\ref{fig:compareEntropy} shows the numbers of correctly labeled voxels and  all observed voxels over robot traveled distance, with only occupancy entropy, with only semantic entropy and with combined entropy.

As shown in the plot, when the observed region is relatively small in an early stage, the benefit of semantic entropy is significant, due to the better exploitation of partial scanned scene. When the robot travels a larger distance, the occupancy entropy starts to take effect, which leads to faster discovery of unknown space and more voxels are observed. Since the semantic entropy has no ability to guarantee discovering more regions, the robot sometimes get stuck in scanned regions with only semantic entropy. With occupancy entropy only, the robot is faster to find new regions such that the total scanned voxels are always the highest. However, when the unknown space is small at a later state, the robot has difficulty to find better observations, which leads to poor performance of semantic segmentation. 
The combined entropy gets the best performance in final scanning results, which works well on both exploration and exploitation jobs. In Figure~\ref{fig:entropyQuality}, we compare semantic segmentation quality of these three different entropy. The visual results verify the above analysis.

\paragraph{Effect of viewing score field}
In the following, we verify the efficiency of our viewing score field-guided path planning approach. We conduct a number of experiments in four synthetic scenes and compare our method with classical path planning algorithm of Dijkstra. Table~\ref{tab} reports total scanning time, traveling distance for our field-guided and Dijkstra path planning on these scenes. The termination conditions are set the same for both algorithms. Here we can see that field-guided panning can save more scanning efforts. To better demonstrate the superiority of our field-guided approach, we also plot the cost time and traveled distance over the number of correctly labeled voxels; see Figure~\ref{fig:comaprePath}. We can observer that the field-guided path planning leads to faster scanning time and less traveled distance all the way.

\begin{table}[t]
	\caption{Comparison between our field-guided method and Dijstkra method. Our field-guided method is much more efficient than the Dijstkra algorithm considering both time cost or explored distance metrics. }
	\centering
\scalebox{0.85}{
\setlength{\tabcolsep}{1.4mm}{
  \begin{tabular}{|c|c|c|c|c|c|}
\hline
\multirow{2}{*}{Scene} & \multirow{2}{*}{Area (m$^2$)} & \multicolumn{2}{c|}{Field-guided method} & \multicolumn{2}{c|}{Dijstkra method}  \\
\cline{3-6}
                       &                                 & Time (s) & Distance(m)                    & Time (s) & Distance (m)                 \\
\hline\hline
1                      & 134.2                           & {\textbf {1237}}    & {\textbf {169.7}}                          & 1412    & 195.1                       \\
\hline
2                      & 239.1                           & {\textbf {1480}}    & {\textbf {195.9}}                          & 1671    & 215.4                       \\
\hline
3                      & 106.9                           & {\textbf {708} }    & {\textbf {92.9} }                          & 1041    & 124.4                       \\
\hline
4                      & 129.5                           & {\textbf {1057}}   & {\textbf {141.1} }                         & 1289    & 171.3                       \\
\hline
Average                & 152.3                           & {\textbf {1121}}    & {\textbf {150.0} }                         & 1353    & 176.5                       \\
\hline
\end{tabular}
}}
  \label{tab}
\end{table}

In addition to the above results, we show more visual results of our active scene understanding in Figure~\ref{fig:scanProcess}. In these examples, it is clear that our scene understanding is guided by collecting more semantic information. Our method would try to drive the robot discover the most semantic objects in a local area before it enters a new area which would maximize the scene understanding efficiency. 




\section{Conclusions}
We have presented a method for active scene understanding based on online RGBD reconstruction
with volumetric segmentation.
Our method leverages the online reconstructed TSDF volume and learns a deep neural network for voxel-based semantic labeling.

It attains the following key features:
\emph{First}, the online scene segmentation is conducted over the online reconstruction, thus benefiting from the
3D spatial reasoning.
\emph{Second}, the robot scanning is guided by the information gain of both geometric reconstruction and semantic understanding.
\emph{Third}, the online estimated viewing score field (VSF) facilitates the joint optimization of both moving path and camera orientation.

We also believe that this work will open up new possibilities for future research:
\emph{First}, our NBV prediction is based on the VSF estimated online. A more favorable approach would be training
a network to achieve an end-to-end NBV estimation. The difficulty lies in how to consider the uncertainty in both reconstruction and segmentation within one neural network.
\emph{Second}, we would like to explore the use of the proposed framework on a real robot.
\emph{Third}, our VSF-based path/trajectory optimization can be extended to support more flexible scanning setting, for example, a robot holding a depth camera in its arm, similar to~\cite{xu2017autonomous}.
\emph{Last}, another interesting future direction would be extending our framework to achieve multi-robot collaborative scene understanding. 

\section{Acknowledgements}
We thank the anonymous reviewers for the valuable comments. This work was supported in part by NSFC (61572507, 61532003, 61622212), Natural Science Foundation of Hunan Province for Distinguished Young Scientists (2017JJ1002), LHTD (20170003), Guangdong Science and Technology Program (2015A030312015) and Shenzhen Innovation Program (KQJSCX20170727101233642).

\bibliographystyle{eg-alpha-doi}
\bibliography{semnbv}

\end{document}